\documentclass[twocolumn,showpacs,preprintnumbers,amsmath,amssymb]{revtex4}
\usepackage{amsmath}

\usepackage{graphicx}%Include figure files
\usepackage{dcolumn}%Align table columns on decimal point
\usepackage{bm}
\usepackage{overpic}

\begin{document}
\title{The Scaling laws of Spatial Structure in Social Networks}
\author{Yanqing Hu\footnote{yanqing.hu.sc@gmail.com}, Yougui Wang, Zengru Di\footnote{zdi@bnu.edu.com}}
 \affiliation{Department of Systems Science, School of Management and Center for Complexity
 Research, Beijing Normal University, Beijing 100875, China}

\date{\today}%It is always \today, today,
             %but any date may be explicitly specified

\begin{abstract}

Social network structure is very important for understanding human
information diffusing, cooperating and competing patterns. It can
bring us with some deep insights about how people affect each other.
As a part of complex networks, social networks have been studied
extensively. Many important universal properties with which we are
quite familiar have been recovered, such as scale free degree
distribution, small world, community structure, self-similarity and
navigability. According to some empirical investigations, we
conclude that our social network also possesses another important
universal property. The spatial structure of social network is scale
invariable. The distribution of geographic distance between
friendship is about $Pr(d)\propto d^{-1}$ which is harmonious with
navigability. More importantly, from the perspective of searching
information, this kind of property can benefit individuals most.

%% add self-similarity

\end{abstract}


 %%%% check!!!!

\maketitle

What does our social network structure look like? How does the
structure benefit us? Understanding the structure of the social
network which has been weaving by us and we live in is a very
interesting problem. As a part of the recent surge of interest in
networks, there has been many researches about social network
\cite{experiment,Lai Ying-Cheng,BA
model,Newman-review,Givens-Newman, Song fractal 1,Song fractal
2,Kleinberg-covergence,new experiment,first issue b, navigation
brief nature,News,Use Kleinberg search,Renaud Lambiotte, full model
navigation}. Social network is a typically complex network. It
possesses some familiar properties such as small-world \cite{first
issue b}, scale free \cite{Newman-review}, community structure
\cite{Givens-Newman} and self-similarity \cite{Song fractal 1, Song
fractal 2}. More interesting, social network has a special property
of navigability \cite{the oldest experiment,new experiment}. The
navigable property of social network has become the subject of both
experimental and theoretical research\cite{new experiment,navigation
brief nature,navigation full,full model navigation,first issue
b,power-law networks Kleinberg search,power-law networks
search,News,Use Kleinberg search,Analyzing kleinberg,Kleinberg
hierarchical model,Oskar licentiate thesis,Renaud Lambiotte,the
oldest experiment}. Recently, Liben-Nowell \textit{et al.} explored
the role of geography alone in routing within a large, online social
network. They used data from about 500 thousand members of the
LiveJournal online community, who made available their state and
city of residence, as well as a list of other LiveJournal friends.
Message-forwarding simulations based on these data showed that a
routing strategy based solely on geography could successfully find
short chains in the network. They also found that the density
function $Pr(d)$ of geographic distance $d$ between friendship is
$Pr(d)\propto d^{-1}$. This result seems contradicted with
Kleinberg's theoretic results \cite{Use Kleinberg search}, which
means our social network is not navigable. Liben-Nowell \textit{et
al.} argued that this kind of contradiction is caused by the
nonuniform population density, then they presented a new model to
explain navigable property of social network. Almost at the same
time, however Lada Adamic and Eytan Ada also found the same
phenomena\cite{power-law networks Kleinberg search}. They
investigated a relatively small social network, the HP email
network. The email network is based on HP buildings. Lada Adamic and
Eytan Ada also cannot explain the contradiction well. They thought
it is caused by the limiting geometry of the buildings. But more
recently, R. Lambiotte \textit{et al.} investigated a large mobile
phone communication network \cite{Renaud Lambiotte}. The network
consists of 2.5 million mobile phone customers that have placed 810
million communications and for whom they have geographical home
localization information. Their empirical result shows that the
mobile phone communication network is corresponding to Kleinberg's
theory. Do Lada Admaic, Eytan Ada,  Liben-Nowell and R. Lambiotte
\textit{et al.} show us a universal phenomenon or just a
coincidence? In this letter we will show that with the distribution
of geographic distance between friendship is $Pr(d)\propto d^{-1}$,
our social network is navigable, even the population density is
nonuniform or some geometry limiting. This kind of distribution is
also harmonic with Kleingber's theory when the density of population
is uniform. So we think this kind of scale invariant distribution of
geographic distance between friendships is universal.

Why does the spatial structure of our social networks possess the
property and how does this distribution benefit us? Even the scaling
law in the spatial structure makes the social network navigable. We
do not think to let the individuals sending message efficiently is
the right answer. In the following two sections we will firstly
conclude that our social network possess the property of
$Pr(d)\propto d^{-1}$ and then we will give the answer to the above
two questions from the perspective of optimal collecting
information.

\section{Spatial Structure and Navigability}
According to the facts mentioned above, we use a scale invariant
friendship network (SIF for a short) \cite{SIFstructure} to model
real social network, even when the population density is nonuniform.
Like Kleinberg's model (K for a short) \cite{full model navigation},
we also employ a lattice as the ground regular network in which each
node possess a weight (population density). Each node $u$ has a
short-range connection to all nodes within $p$ lattice steps, and
$q$ long-range connections generated independently from a
distribution $Pr(d)\sim d^{a}$ (density function). In order to keep
model simple and not to lose any generality \cite{full model
navigation}, we always set $q=p=1$. For each long-range connection
of $u$, we first randomly choose a distance $d$ according to the
above power law distribution. Then randomly choose a node $v$
(proportional to $v$'s weight) from the node set in which the
distance for $u$ to each element is $d$. At last, generate a
directed long-range connection from $u$ to $v$. When population
density is nonuniform, compared with K model, SIF always keeps the
distribution of geographic distance between friendship scale
invariant in any situations. When the population density is uniform,
the probability that node $u$ chooses node $v$ as its long-range
contact in SIF is
\begin{equation}
Pr_{SIF}(u,v,a)=\frac{1}{c(u,v)}\frac{d(u,v)^{a}}{\sum_{d=1}^{L}d^{a}}
\end{equation}
where $c(u,v)=|\{x|d(u,x)=d(u,v), x\in S\}|$, $S$ is the set of all
nodes in SIF network and
\begin{equation}Pr_{K}(u,v,\beta)=\frac{d(u,v)^{\beta}}{\sum_{x\neq
u}d(u,x)^{\beta}}\end{equation} in K network \cite{full model
navigation}. We have
\begin{equation}\frac{Pr_{K}(u,v,-k)}{Pr_{SIF}(u,v,-1)}=1\end{equation} for
$k$-dimensional lattice based network. It implies that SIF network
with $a=-1$ corresponds to the result in K network with $\beta=-k$
when the population density is uniform. Here, we should note that in
$k$-dimensional based lattice, if node $u$ connect to node $v$ with
probability proportional to $d(u,v)^{\beta}$, it does not equal
$Pr(d)\propto d^{\beta}$, but $Pr(d)\propto d^{k+\beta-1}$. From the
above discussion, we know that Kleinberg's result is not
contradicted with the empirical results but well correspond to them.

We also can prove that, our social network is navigable just
according to the distribution of geographic distance between
friendship is $Pr(d)\propto d^{-1}$. The expectation of
decentralized search is at most $O(log^{4}n)$ for nonuniform
population density. Here, we focus on the 2-dimensional lattice, and
the analysis can be applied to $k$-dimensional lattice networks. We
can easily make the following two assumptions (1) In each small
enough region, the population density is uniform. (2) The maximum
population of all small region are $M$ and minimum population is
$m>0$. Under the two assumptions, easily we have
\begin{equation}
c\frac{M}{m}d^{-1}\leq Pr(d) \leq c\frac{m}{M}d^{-1}\label{KSIF}
\end{equation}
in K network with $\beta=-2$, where $c$ is a constant.

Starting from the analysis of time complexity of navigation, we
compare the following two routing strategies. Strategy $A$, the
message that navigates to target $t$ directly (the original
Kleinberg's greedy routing). Strategy $B$, the message firstly
navigates to node $j$, then from node $j$ navigates to target $t$.
Obviously, the expectant steps spent by Strategy $B$ is not less
than Strategy $A$ for any node $j$. Thus we have the expectant steps
spent on navigation in any small region is at most $O(\log^2 n)$.
Each small region can be regarded as a node, then we get a new
2-dimensional lattice in which each node's weight (population) is
between $M$ and $m$. According to the article \cite{navigation
full}, we have the expectant steps spent among the squares is at
most $O(\frac{M}{m}\log^2 n)$. Thus, the upper bound of navigation
in nonuniform lattice is $O(\frac{M}{m}\log^4 n)$. So, with the
above spatial structure property, our social network is navigable.

\begin{figure}
\center
\includegraphics[width=7cm]{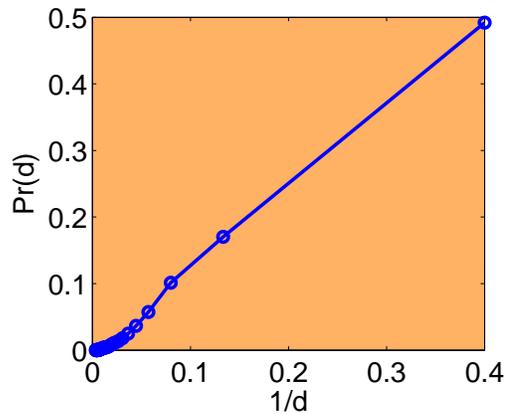}
\caption{The spatial structure in mobile communication network. From
it we can see that $Pr(d)$ and $\frac{1}{d}$ have a roughly linear
relationship. You can get more detail about the data from the
article \cite{Renaud Lambiotte}.}\label{Empirical_data}
\end{figure}

From Eq. \ref{KSIF}, we can see that if the difference of population
density among different areas are not too big, SIF with $a=-1$ and K
model with $\beta=-2$ for 2-dimensional situation have no essential
difference. R. Lambiotte provided us the data freely. From their
data we also can see the same phenomenon showed by Lada Admaic,
Eytan Ada and Liben-Nowell \textit{et al.}. Fig.
\ref{Empirical_data} presents the relationship between $Pr(d)$ and
$\frac{1}{d}$, we can see that they have a linear relationship
roughly. So from the above empirical investigations and theoretical
discussion we can certainly draw a conclusion that distribution of
geographic distance between friendships is
\begin{equation}
Pr(d)\propto d^{-1}\end{equation}.

\section{Optimal Collecting Information}

Now we face some questions, why does this kind of distribution exist
in social networks and how does it benefit individuals? In our
social networks, many human economical behaviors can be roughly
regarded as collecting information. Making friends also can be looked
as the way to search information. So, the social network should be
an optimal network which can benefit people for collecting information.
What is the optimal $a$ in social network? The following model will
give us a possible answer.

Suppose, individuals have average finite energy $w$ which can be
represented by the sum of distances between one and his/her friends.
For a node $u$, each time, as the rule of SIF, we first randomly
choose a distance $d$ according to  $Pr(d)\propto d^{a}$, then
randomly choose a node $v$ for the node set in which the distance
for $u$ to each element is $d$. The information that node $v$ bring
to $u$ can be denoted by node $v$ and all nodes within $p$ lattice
steps to node $v$. After a proper time, the sum of all $d$ chosen
will approach $w$, then stop the execution. The information that $u$
hunted can be expressed by the sequence of nodes. We use the entropy
of the sequence to denote the value of information. Then we have the
model: $\max\varepsilon=-\sum^{n}_{i=1}p_i\ln{p_i}$ subjected to
$\sum_{j=1}^{m}d(u,j)=w$ and $Pr(d)\propto d^{a}$, where, $p_i$
denotes the frequency of node $i$ in the information sequence. For
instance, if the information sequence is $\{1, 1, 2, 3, 7\}$, then
$p_1=\frac{2}{5}$ and $p_2=p_3=p_7=\frac{1}{5}$, others are 0.
Obviously, more large the $\varepsilon$, more information hunted.
Here, we let $p=1$ (if $p$ is not too large we can get the same
result). The reason is that, according to our common sense, if $B$
is a friend of $A$, $A$ will know more information about people who
are always around $B$. Actually, we should take account all friends
of $B$, but the time complexity will be expensive. We compare the
two kinds of simulations in not too large networks, they have the
similar results.

\begin{figure}
\center
\includegraphics[width=4.5cm,height=3.6cm]{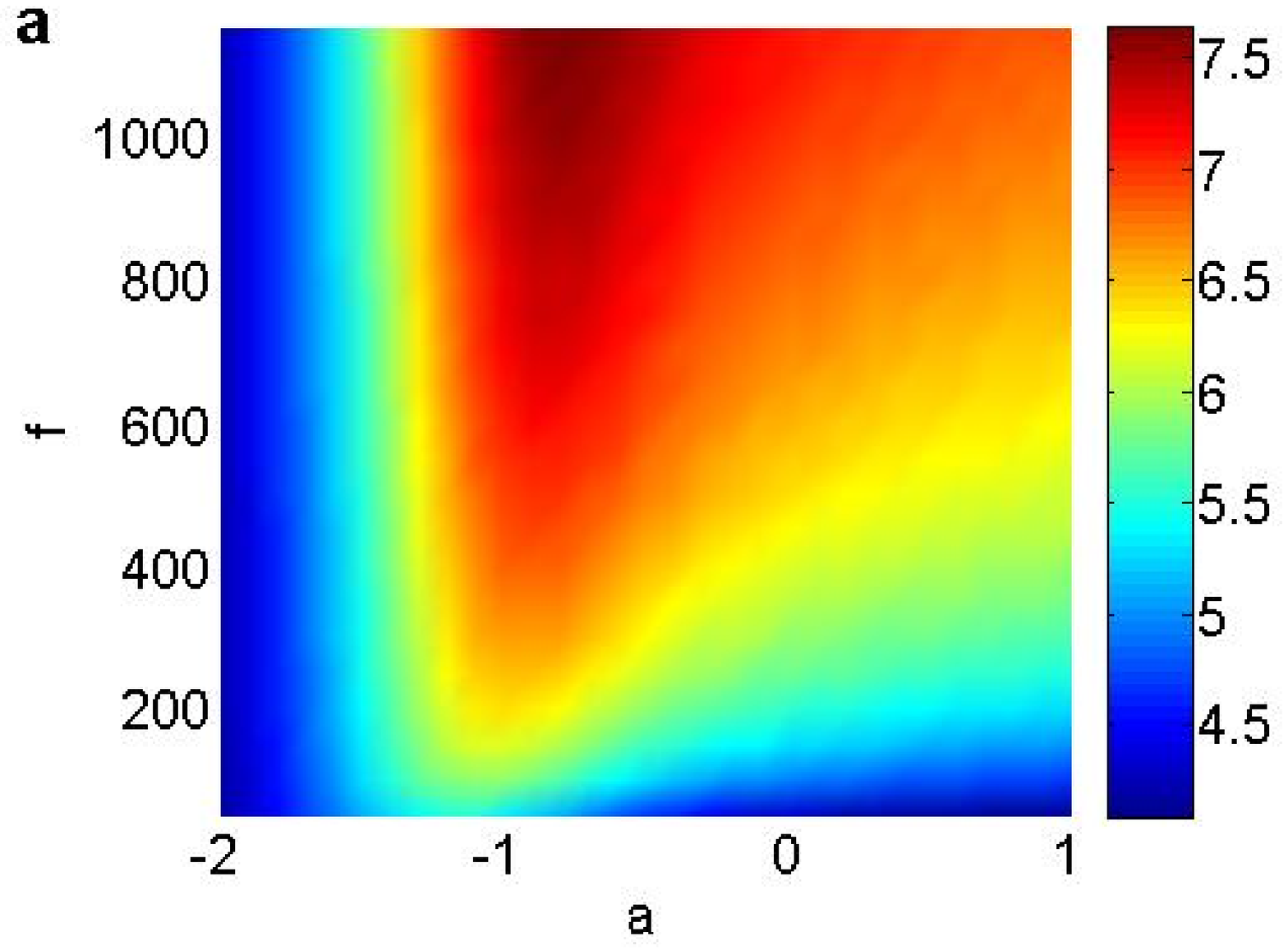}\includegraphics[width=4.0cm,height=3.5cm]{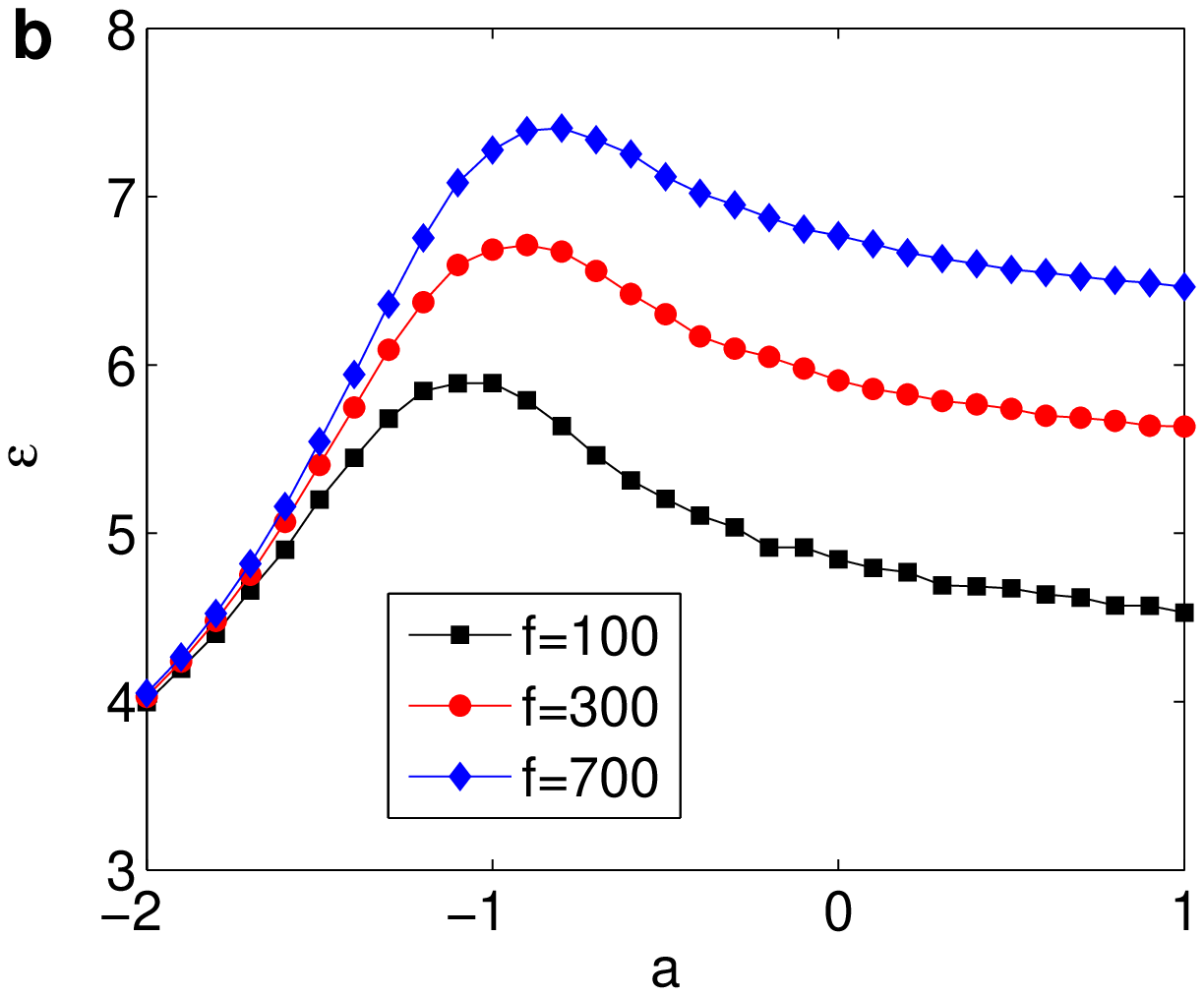}
\includegraphics[width=4.3cm,height=3.5cm]{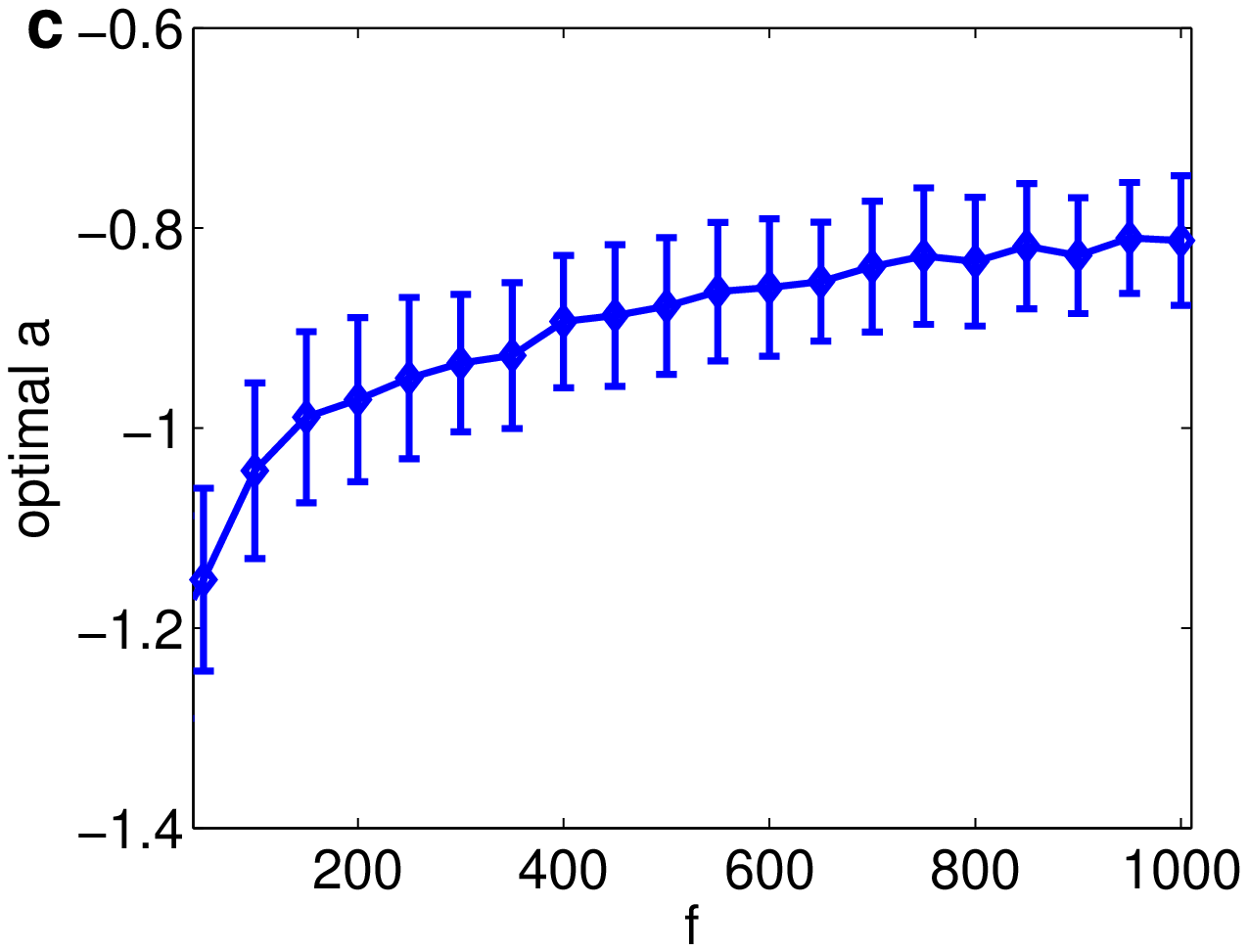}\includegraphics[width=4.3cm,height=3.5cm]{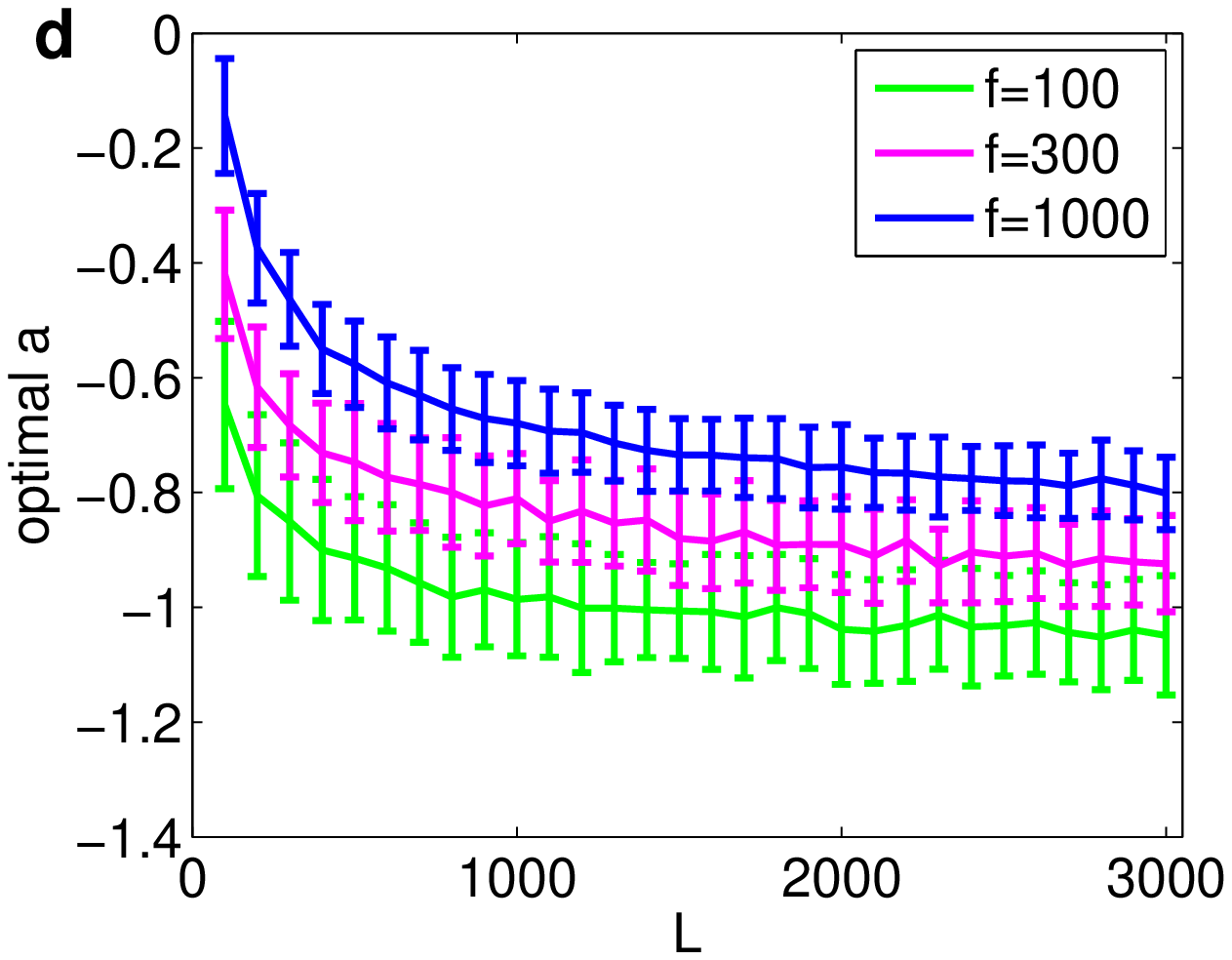}
\caption{The relationship among $\varepsilon, f$, $a$ and $L$.
\textbf{a}, we shows the relationships among $f$, $a$ and
$\varepsilon$, where $L=3600$. We use color to indicate the value of
$\varepsilon$. \textbf{b}, depicts the changes of information
entropy $\varepsilon$ with the changes of $a$ when $f=100, 300,
700$. \textbf{c}, shows the optimal $a$ with the changes of average
friends number $f$. The error bars denotes the standard deviations.
\textbf{d}, shows the relationships between optimal $a$ and the
lattice size $L$. The error bars denotes the standard deviations.}
\label{optimal}
\end{figure}

We simulate the above model on a toroidal lattice. The largest
distance among pair of nodes in the lattice is $L=3600$. For
America, from the north to south and from the west to east the
largest distances are $4500km$, and $2700km$ respectively, and the
average is about $3600km$ (here, it is no necessary to make the
parameter so much accuracy in the model). The average number of
friends we contact in one year is about $f=300$
\cite{number_friends}. According to the empirical result
$Pr(d)\propto d^{-1}$, we can calculate the average $w=\frac{f\cdot
L}{\log{L}}$. Note that here the empirical result of $Pr(d)\propto
d^{-1}$ is only used to determine the parameter value of the model.
It is independent of the optimal $a$. Fig. \ref{optimal} shows the
relationship between $a$ and $f$. We can find that, the optimal $a$
depends on $f$. When $f$ is about 300, the optimal $a=-0.94\pm0.08$
($\pm$ standard deviations). This indicates that when people just
posses finite energy, it is a good way to keep friendships holding
$Pr(d)\propto d^{-1}$.

\section {Conclusion}
From the empirical results, we conclude that the distance
distribution between friendship is scale invariant. The
distributions is about $Pr(d)\propto d^{-1}$ which is an important
and universal property for social network. It not only makes our
social network is navigable but most importantly it can benefit
individuals for searching information.

\section*{Acknowledgement}
We appreciate Dr. R. Lambiotte for providing mobile network data.
Yanqing Hu wants to thank Prof. Fukang Fang and Prof. Gang Hu for
some very useful discussions and  Dr. Erbo Zhao for some help in
English writing. The work is partially supported by NSFC under Grant
No. 60534080 and No. 70771011.

\end{document}